\documentclass[preprint]{revtex4-1}
\usepackage{graphicx}
\usepackage{dcolumn}
\usepackage{bm}
\usepackage{graphicx}
\usepackage{amsmath}
\usepackage{graphicx}
\usepackage{float}
\usepackage{amssymb}
\usepackage{datetime}
\usepackage{amsmath}
\usepackage[english]{babel}
\usepackage[utf8]{inputenc}
\usepackage{graphicx}
\usepackage{epstopdf}
\usepackage{amsmath}
\usepackage{blindtext}
\usepackage{subfigure}
\usepackage{caption}

\begin{document}
\title{Generalized lower-hybrid mode with, density gradient, equilibrium ExB drift, collisions and finite electron Larmor radius: Numerical studies with MATLAB solver.}
\author{\firstname{Ivan}~\surname{Romadanov}}
\email{ivr509@mail.usask.ca}
\author{\firstname{Andrei}~\surname{Smolyakov}}
\email{andrei.smolyakov@usask.ca}
\author{\firstname{Winston}~\surname{Frias}}
\email{winston.frias@usask.ca}
\author{\firstname{Oleksandr}~\surname{Chapurin}}
\email{alex.chapurin@usask.ca}
\author{\firstname{Oleksandr}~\surname{Koshkarov}}
\email{olk548@usask.ca}
\affiliation{$^1$Department of Physics and Engineering Physics, University of Saskatchewan, Saskatoon SK S7N 5E2, Canada}

\begin{abstract}
	MATLAB solver for general dispersion relation has been developed to study local instabilities in partially magnetized plasmas typical for $\mathbf{E \times B}$ discharges. Several examples of the Simon-Hoh, lower-hybrid and ion-sound instabilities with the parameters typical for Penning discharge are shown. This solver allows parametrically investigate the local dispersion relation for the wide range of plasma parameters and prepare reports.
\end{abstract}

\maketitle

\section{INTRODUCTION}
\vspace*{-1em}

Solver, "Hall Plasmas Discharge Solver" (or "HPDSolver") has been developed to investigate local instabilities in partially magnetized plasmas typical for $\mathbf{E\times B}$ discharge plasmas such as in Hall thrusters, Penning discharges, and magnetrons. This solver has user interface, which allows easily change various plasma parameters and obtain results. The form of dispersion relation can be changed; thus, different physical effects can be included or excluded and their effects on the unstable modes can be investigated. In this paper, examples for the Simon-Hoh, lower-hybrid, and ion-sound instabilities are presented, together with the description of the code, and its features. Detailed explanations of the physics models, equations and notations are given in the companion paper in Ref. \onlinecite{SmolyakovPPCF2016}. The source code and executable installer can be downloaded by link \onlinecite{Git}. Current version of the installer can be used on the MS Windows 8 and 10 (the only two versions tested so far) computers and does not require the installed MATLAB. 
\vspace*{-1em}

\section{GENERAL DISPERSION RELATION FOR PENNING DISCHARGE}
\vspace*{-1em}

Important part of instability studies is an experimental investigation of plasma fluctuations and validation of theoretical and numerical models. Hall thruster is not convenient from experimental point of view, especially when probe measurements are performed. Thus, more flexible and versatile setup is required. Experiments with Penning discharges have long been demonstrating many instabilities which seem to be similar to those in Hall thrusters. One could expect that there are essential common physics elements between these two systems. The geometry of Penning trap is similar to the geometry of Hall thrusters especially in latest designs with strong axial magnetic field. The particular experimental design at  Princeton Plasma Physics Laboratory\cite{RaitsesIEPC2015} has the axial beam of energetic electrons that serves as a cathode and the coaxial outer electrode as an anode. The applied voltage creates the radial electric field and the axial magnetic field is created by the external solenoid, thus creating crossed E and B field configuration.

Because of above advantages, theoretical investigation of Penning discharge seems to be a reasonable part of general study of $\mathbf{E\times \mathbf{B}}$ discharges, especially for verification of the proposed theories. Therefore, in this paper we use parameters of Penning discharge system to investigate some predictions of local model. Typical parameters of the Penning discharge device\cite{RaitsesIEPC2015} are presented in Table \ref{Table:param}.

\begin{table}[h!]
\caption{Typical Penning discharge parameters}
\vspace*{-1em}
	\begin{center}
	\begin{tabular}{l|c}
		Parameter       & Xe                          \\ \hline
		$T_e$           & 10 $eV$                     \\ \hline
		$B$             & 100 $G$                     \\ \hline
		$E$             & $100$ $V/m$                 \\ \hline
		$n$             & $10^{18}$ $m^{-3}$          \\ \hline
		$L_n$           & -$5\cdot10^{-2}$ $m$ 		  \\ \hline
		$c_s$           & $2.7\cdot 10^3$ $m/s$       \\ \hline
		$v_{Te}$        & $1.3\cdot 10^6$ $m/s$       \\ \hline
		$v_*$           & $1.0\cdot 10^5$ $m/s$       \\ \hline
		$v_0$           & $1.0\cdot 10^4$ $m/s$       \\ \hline
		$\omega_{LH}$   & $3.6\cdot 10^6$ $rad/s$     \\ \hline
		$\omega_{ce}$   & $1.8\cdot 10^9$ $rad/s$     \\ \hline
		$\omega_{ci}$   & $7.3\cdot 10^3$ $rad/s$     \\ \hline
		$\omega_{pe}$   & $5.6\cdot 10^{10}$ $rad/s$  \\ \hline
		$\rho_e$        & $7.5\cdot 10^{-4}$ $m$      \\
	\end{tabular}
	\label{Table:param}
	\end{center}
\end{table}
\vspace*{-2em}

\noindent
where $T_e-$ electron temperature, $B-$ magnetic field, $n - $ density, $E-$ electric field, $L_n-$ density gradient length, $c_s-$ ion sound velocity, $v_s-$ diamagnetic drift velocity, $v_0-$ $\mathbf{E}\times\mathbf{B}$ velocity, $\omega_{LH}-$ lower hybrid frequency, $\rho_e-$ electron gyroradius.

Local models allow to investigate some characteristic features of the plasma discharges without involving computation expensive and time consuming simulations. General dispersion relation, which includes effects of electron-neutral collisions, density gradients, electron inertia and gyroviscosity, was developed \cite{SmolyakovPPCF2016, FriasPoP2013, FriasPoP2014}. For this model, $\mathbf{E}$ is in $x-$ direction, $\mathbf{B}$ is in $z-$ direction; $y$ is a poloidal coordinate:
\begin{eqnarray}
	\dfrac{\omega_*-\omega_D+(\omega - \omega_0+i\nu_{en})k^2_{\perp}\rho^2_e + ik^2_{\parallel}D_{\parallel}}{\omega-\omega_0-\omega_D+(\omega - \omega_0+i\nu_{en})k^2_{\perp}\rho^2_e + ik^2_{\parallel}D_{\parallel}}=\dfrac{k^2_{\perp}c^2_s + k^2_{\parallel}c^2_s}{(\omega - \omega_i)^2}
\end{eqnarray}
where $\omega_*-$ diamagnetic drift frequency, $\omega_0-$ $\mathbf{E\times B}$ drift frequency, $\nu_{en}-$ electron-neutral collision frequency, $\omega_i = k_xV_i$ - ion flow frequency, $D_{\parallel}$ - diffusion coefficient in $z$ direction, $k_{\perp} = k_x^2 + k_y^2$ - wave vector in the plane perpendicular to $\mathbf{B}$, $k_{\parallel}$ - wave vector, parallel to $\mathbf{B}$.
\vspace*{-1em}
\section{MATLAB SOLVER}
\vspace*{-1em}
\subsection{User Interface}
\begin{figure}[h]
	\centering
	\includegraphics[width=8cm]{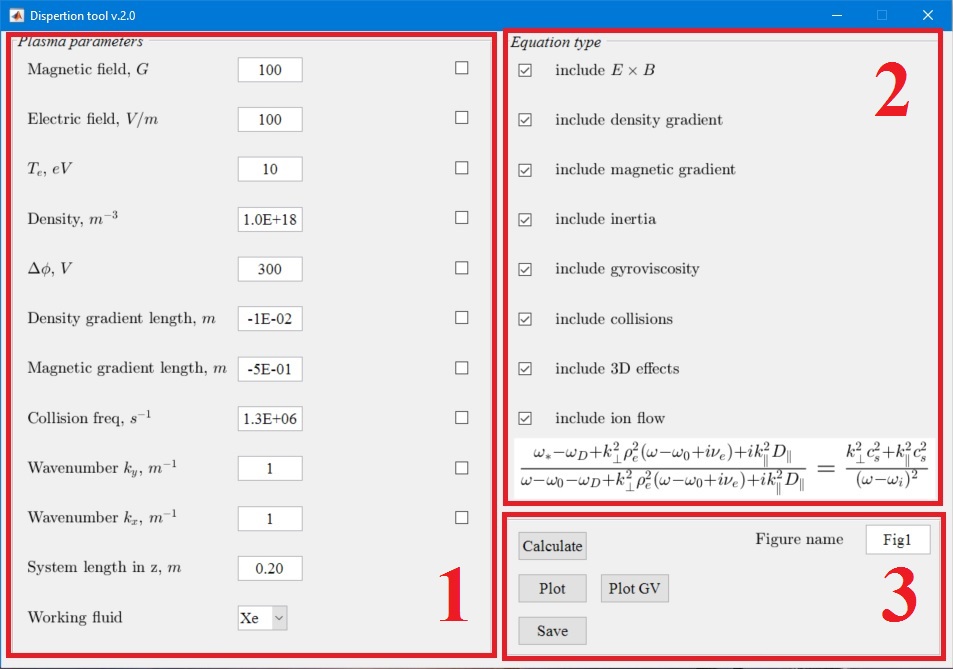}
	\caption{General view of the software interface. Section 1 -- plasma parameters control; section 2 -- dispersion relation form; section 3 -- control buttons.}
	\label{fig:FigA1}
\end{figure}

\vspace*{-1em}
Hall Plasmas Discharge solver can be conveniently used to study parametric dependence of local instabilities for various plasma parameters and different physical effects. General view of the software interface is shown in Fig. \ref{fig:FigA1}. It consists of three main fields: 1 -- plasma parameters; 2 -- type of the solved dispersion relation; 3 -- control buttons to calculate, plot, and save results.

\begin{figure}[h]
	\centering
	\includegraphics[width=6cm]{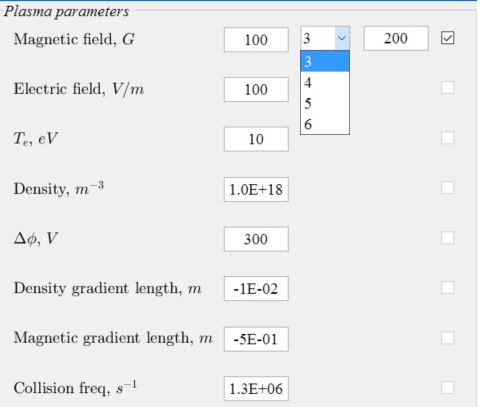}
	\caption{Example of selecting the magnetic field as a variable parameter.}
	\label{fig:FigA2}
\end{figure}

\vspace*{-2em}
At the startup all parameters are field with some initial values, user can change them accordingly. Regular or scientific notation can be used for input values. Checkboxes on the right side allow to set required parameter as a variable. An example for the magnetic field is shown in Fig. \ref{fig:FigA2}. 

\vspace*{-1em}

\vspace*{-2em}
When user selects the checkbox, two additional fields appear: field for the final value of the parameter and field where user can select the number of points within the interval. It is important to notice that all other checkboxes, except for $k_x$ and $k_y$ become disabled; thus, user is prevented from selecting other plasma parameters as variables. System length cannot be selected as a variable parameter.

\begin{figure}[h]
	\centering
	\includegraphics[width=6cm]{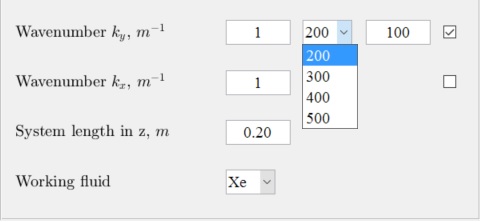}
	\caption{Example of selecting the $k_y$ wavenumber as a variable parameter.}
	\label{fig:FigA3}
\end{figure}

An example of setting $k_y$ wavenumber as a variable is shown in Fig. \ref{fig:FigA3}. Meaning of fields is the same, but the number of point within the interval is bigger, thus a smooth dependency can be obtained. If both $k_x$ and $k_y$ are selected, all checkboxes for plasma parameters will become disabled, thus user cannot set anything else as a variable. 

Either $k_x$ or $k_y$ wavenumber is the mandatory parameter to select. In case a user hits the "Calculate" button without selection of $k_x$ or $k_y$ a warning sign will appear. Therefore, the next options for the output are available: a single 2D graph or a set of 2D graphs (in case one of the plasma parameter is chosen as a variable), which represent the dependency of real and imaginary part of $\omega$ as a function of wavenumber (in $x$ or $y$ direction) at chosen set of plasma parameters; another option is two 3D surfaces, which represent dependency of real and imaginary part of $\omega$ as a function of both wavenumbers in $x$ and $y$ directions. 

Checkboxes in section 2 are used to control the form of the solved dispersion relation. Corresponding equation is shown at the bottom of the section 2, and it will be changed according to the selected parameters. One has to remember, that it is possible to modify the dispersion relation in a way that no instability will occur.

\begin{figure}[h]
	\centering
	\includegraphics[width=6cm]{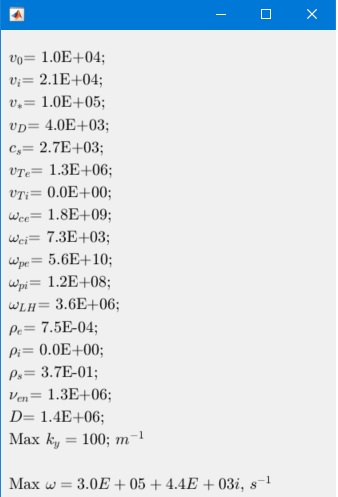}
	\caption{Plasma parameter window.}
	\label{fig:FigA4}
\end{figure}

Control button are grouped in section 3. Once "Calculate" button is pressed, the pop-up window with process bar will appear. After calculations are completed the window with plasma parameters will appear. It shows all velocities, plasma frequencies, and maximum values of $\omega$ together with corresponding wavenumber, at which this frequency is reached. Example is presented in Fig. \ref{fig:FigA4}.

\newpage
Plot functions are executed by "Plot" button. As it was mentioned before, it is possible to plot 2D and 3D graphs, depending on chosen parameters. "PlotGV" button plot group velocity for the unstable modes. Output can be in 2D and 3D form as well. Group velocity is calculated numerically by
\begin{eqnarray}
	v = \dfrac{\partial\omega}{\partial k}
\end{eqnarray}

\begin{figure}[h]
	\centering
	\includegraphics[width=6cm]{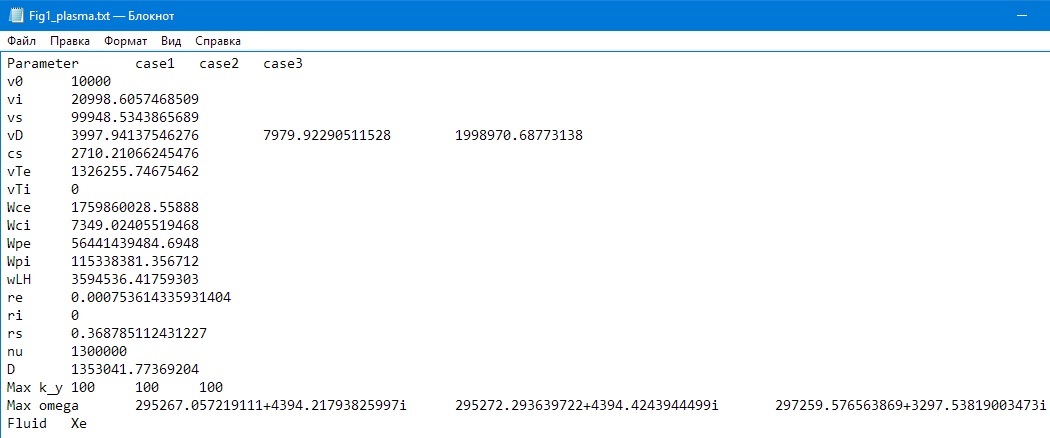}
	\caption{Plasma parameter window.}
	\label{fig:FigA5}
\end{figure}

Save button generates two files. First file named as "Figure name" field + "\_plasma.txt", contains calculation parameter fields and plasma parameters. It is a tab delimiter file, which later can be read by user from other applications. Files are saved in the same directory where executable file is located. Example of the structure of this file is shown in Fig. \ref{fig:FigA5}.

First column is the parameter name, second is the value. If one of the parameter was chosen as a variable, additional columns are created. Maximum values of omegas and corresponding wavenumbers are saved as well.

Second file is named as "Figure name" field + "\_data.txt", it contains calculation results, which is organized in columns: "Parameter", "KY", "KX", "ROOT1",	"ROOT2", "ROOT3". Parameters column is the string which contains the name of the field which was hose as a variable. If no field was chosen then it is empty. "KY", "KX" are wavenumbers. "ROOT" columns all roots for the dispersion relation. There can be 2 or 3 roots, depending on the form of the equation.
\subsection{Installation}
\vspace*{-1em}
To install the software, one has to download "HPDSolver\_web.exe" by link \onlinecite{Git}. During the installation it will automatically download and install all necessary files. However, if you already have MATLAB run-time engine, you can download only "HPDSolver.exe", "constants.mat", and "splash.png" files and run it directly. Remember, that "constants.mat" is required to run the solver, otherwise it will crash.

\section{EXAMPLES OF INSTABILITIES FROM GENERAL DISPERSION RELATION}

Here we present several cases of the unstable modes which can be found with the solver. All used parameters are taken from Table \ref{Table:param}.

\subsection{Simon-Hoh instability}
When the stationary electron current due to the $\mathbf{E\times B}$ drift is present, the anti-drift mode becomes unstable due to the phase shift between the potential and electron density. Resulting gradient drift instability is described by the dispersion equation
\begin{equation}
-\dfrac{\omega_{ci}k_{\perp }^{2}L_{n}}{k_{y}}=\dfrac{\omega ^{2}}{\omega-\omega _{0}},  \label{eq:local_SH}
\end{equation}
where $\omega _{0}=\mathbf{k\cdot V}_{E}$ is the azimuthal (closed drift) flow of electrons in crossed $\mathbf{E}\times \mathbf{B}$ fields and $\mathbf{E}_{0}=E_{0}\widehat{\mathbf{x}}$. This is the reactive instability of negative energy perturbations with a phase velocity below the stationary $V_{E}=\mathbf{E\times B}/B^{2}$ velocity. This mode is referred here as the collisionless Simon-Hoh instability \cite{SimonPF1963, HohPF1963, TaoPF1993}. The condition $\mathbf{E\cdot \nabla }n>0$ is required for the instability \cite{SakawaESHB}. From Eq. (\ref{eq:local_SH}) one gets:
\begin{equation}
\omega _{0}\omega _{\ast }>\dfrac{1}{4}k_{\bot }^{2}c_{s}^{2}.
\label{eq:local_crit}
\end{equation}

\begin{figure}[h]
	\centering
	\includegraphics[width=6cm]{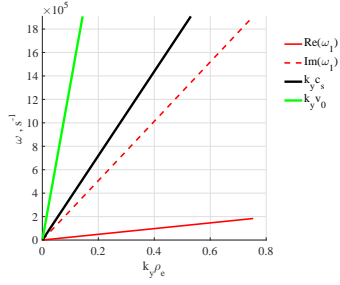}
	\caption{Growth rate and frequency of the Simon-Hoh mode as a function of $k_{y}\rho _{e}$, and $k_x = 1$ $m^{-1}$.}
	\label{fig:local_ky}
\end{figure}

For given parameters, dependency of growth rate and frequency on $k_{y}\rho _{e}$ are presented in Fig. \ref{fig:local_ky}.
\vspace*{-1em}

\begin{figure}[h]
	\centering
	\includegraphics[width=6cm]{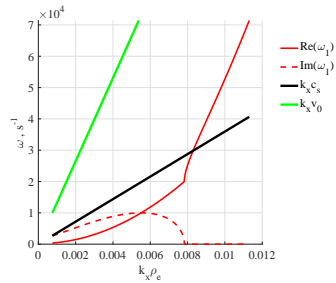}
	\caption{Growth rate and frequency of the Simon-Hoh mode as a function of $k_{x}\rho _{e}$, and $k_y = 1$ $m^{-1}$.}
	\label{fig:local_kx}
\end{figure}

The instability condition is most easily satisfied for $k_{x}\rightarrow 0$. However, the modes with low values of $k_{x}$ are not the fastest growing modes. In general, there exist multiple eigenmodes with different values of $k_{x};$ roughly, the value of $k_{x}$ may be associated with a number of nodes in the $x$ direction. For a given $k_{y}$, the dependence of the growth rate on $k_{x}$ is nonmonotonous, see Fig. \ref{fig:local_kx}.

\vspace*{-1em}

It follows from (\ref{eq:local_crit}) that for high values of $k_{x}$, the mode becomes stable. The maximum value of $k_{x}$ for unstable modes increases with $k_{y}$. The growth rate is maximized when the following condition is satisfied
\begin{equation}
k_{y}^{2}+k_{x}^{2}=2\dfrac{\omega _{\ast }\omega _{0}}{c_{s}^{2}}=2\dfrac{%
	E_{0}}{B_{0}}\dfrac{k_{y}^{2}}{\omega _{ci}L_{n}}.  \label{eq:max_kx}
\end{equation}

Using (\ref{eq:max_kx}) one finds from (\ref{eq:local_SH}) that the modes with the largest growth rate have the local eigenmode frequency given by the expression
\begin{equation}
\omega =\omega _{0} + i\omega _{0}.
\label{eq:max_om}
\end{equation}

Therefore, for any given azimuthal wavenumber, $k_{y}$, the real part of the frequency and growth rate for the most unstable local mode are equal to the $\mathbf{E\times B}$ frequency.

\begin{figure}[h!]
	\centering
	\includegraphics[width=8cm]{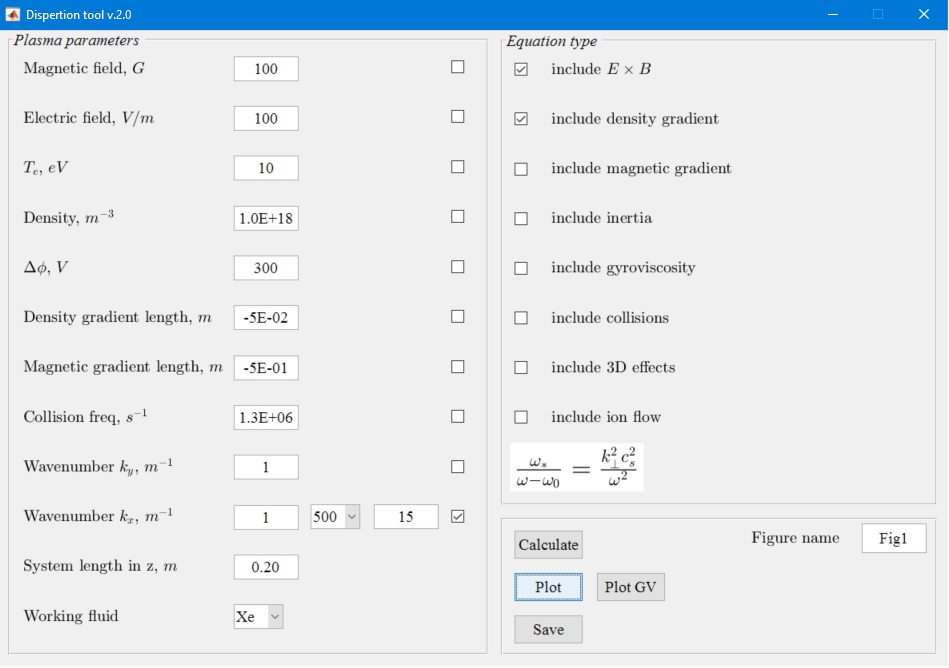}
	\caption{View of the solver interface with parameters, which were used to
		obtain Fig. \ref{fig:local_ky}, \ref{fig:local_kx}.}
	\label{fig:interface_SH}
\end{figure}

Solver interface with parameters, which were used to calculate data for Fig. \ref{fig:local_ky}, \ref{fig:local_kx}, is shown in Fig. \ref{fig:interface_SH}. As it is seen, interface allows to control the form of the equation, what makes it more convenient.

\subsection{Effect of electron inertia on Simon-Hoh instability}

As was noted in \cite{SmolyakovPPCF2016}, Simon-Hoh instability is, in fact, can have high frequency comparable to $\omega _{LH}$ even for low $k_{y}$. Therefore, electron inertia effect become important and should be included. Corresponding dispersion relation is given in Eq. \ref{eq:Eq2}

\begin{equation}
\dfrac{\omega _{\ast }+k_{\perp }^{2}\rho _{e}^{2}(\omega -\omega _{0})}{%
	\omega -\omega _{0}}=\dfrac{k_{\perp }^{2}c_{s}^{2}}{\omega ^{2}}.
\label{eq:Eq2}
\end{equation}

\begin{figure}[h!]
	\centering
	\includegraphics[width=6cm]{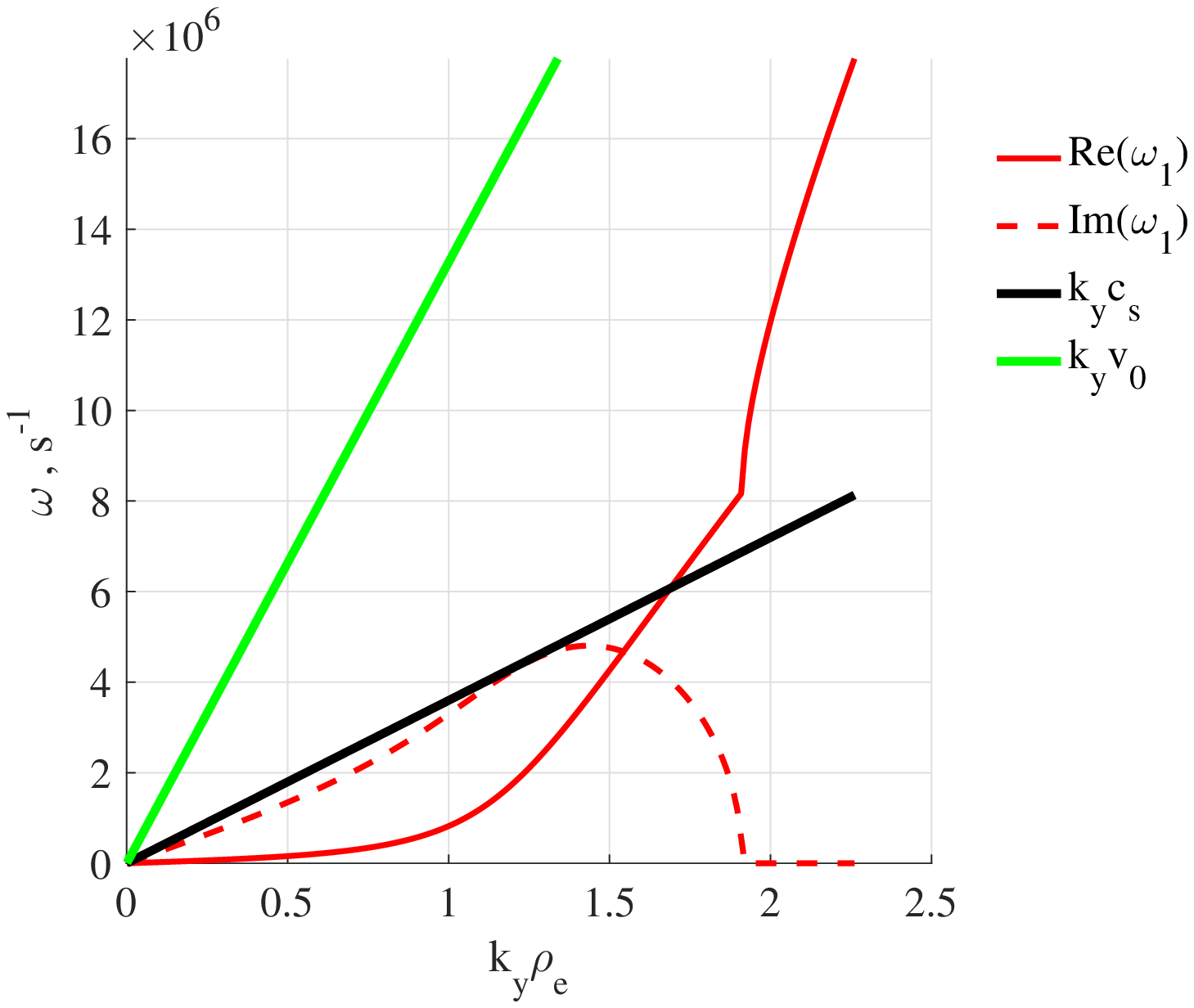}
	\includegraphics[width=6cm]{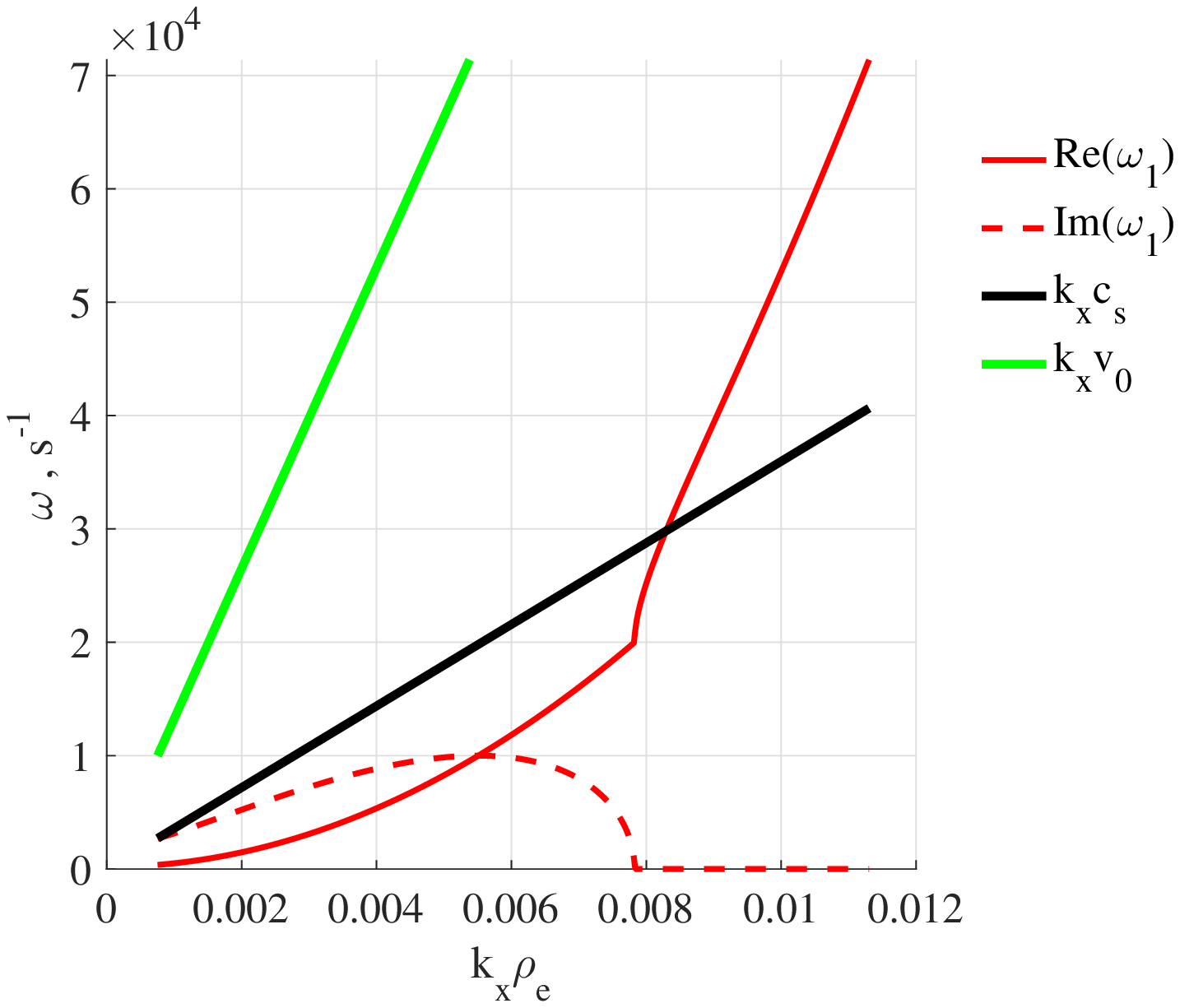}
	\caption{Frequency and growth rate of the Simon-Hoh instability with electron inertia effect as functions of $k_y\protect\rho_e$ at $k_x = 1$ (left) and $k_x\protect\rho_e$ at $k_y = 1$ (right).}
	\label{fig:inertia}
\end{figure}

Typical mode behavior is shown in Fig. \ref{fig:inertia}. Electron inertia results in the cut-off for the instability at high $k_{y}$ values, while dependence on $k_{x}$ remains approximately the same. Note that the eigenmode frequencies become lower, compared to the Simon-Hoh case.

\begin{figure}[h!]
	\centering
	\includegraphics[width=6cm]{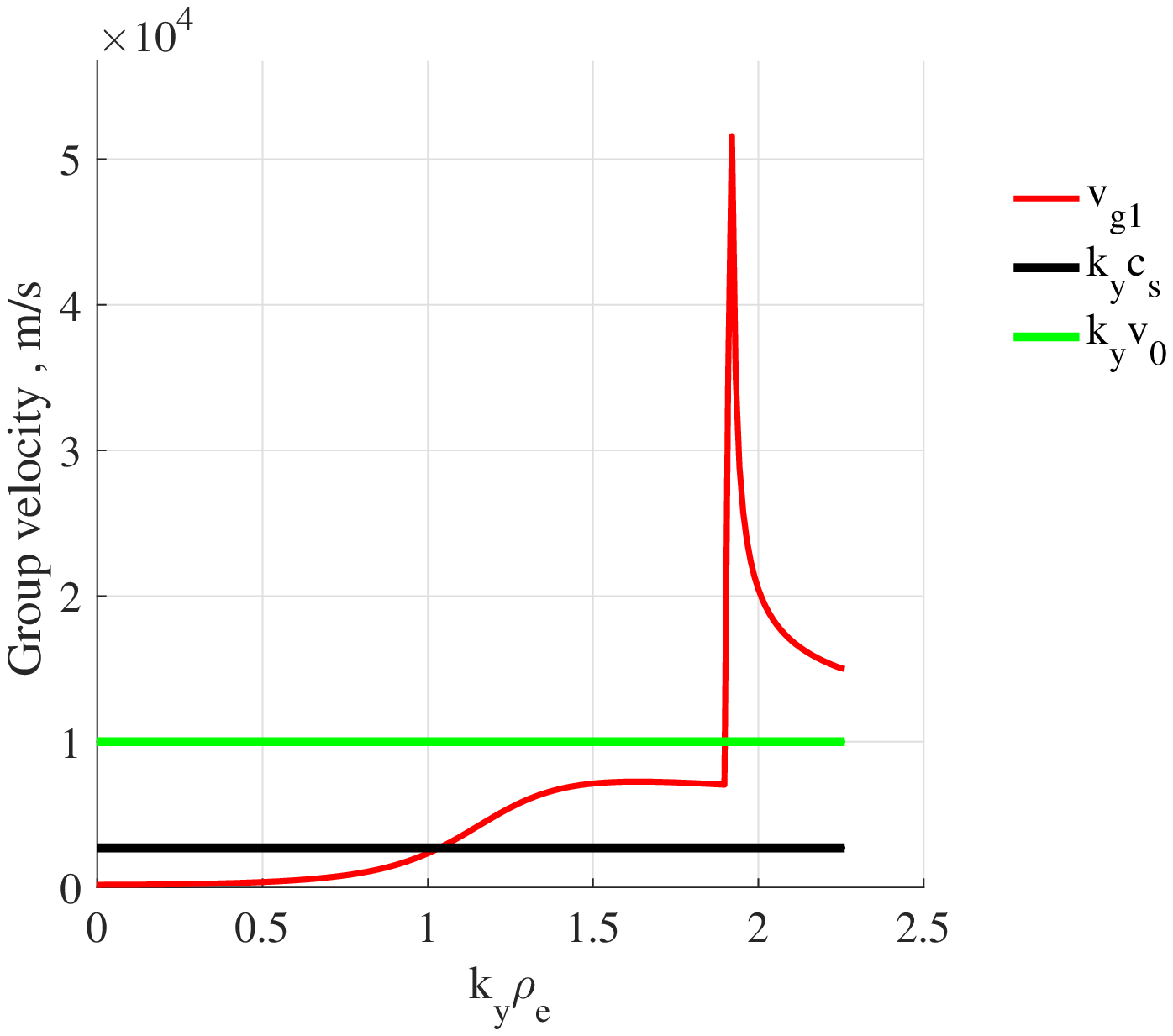}
	\includegraphics[width=6cm]{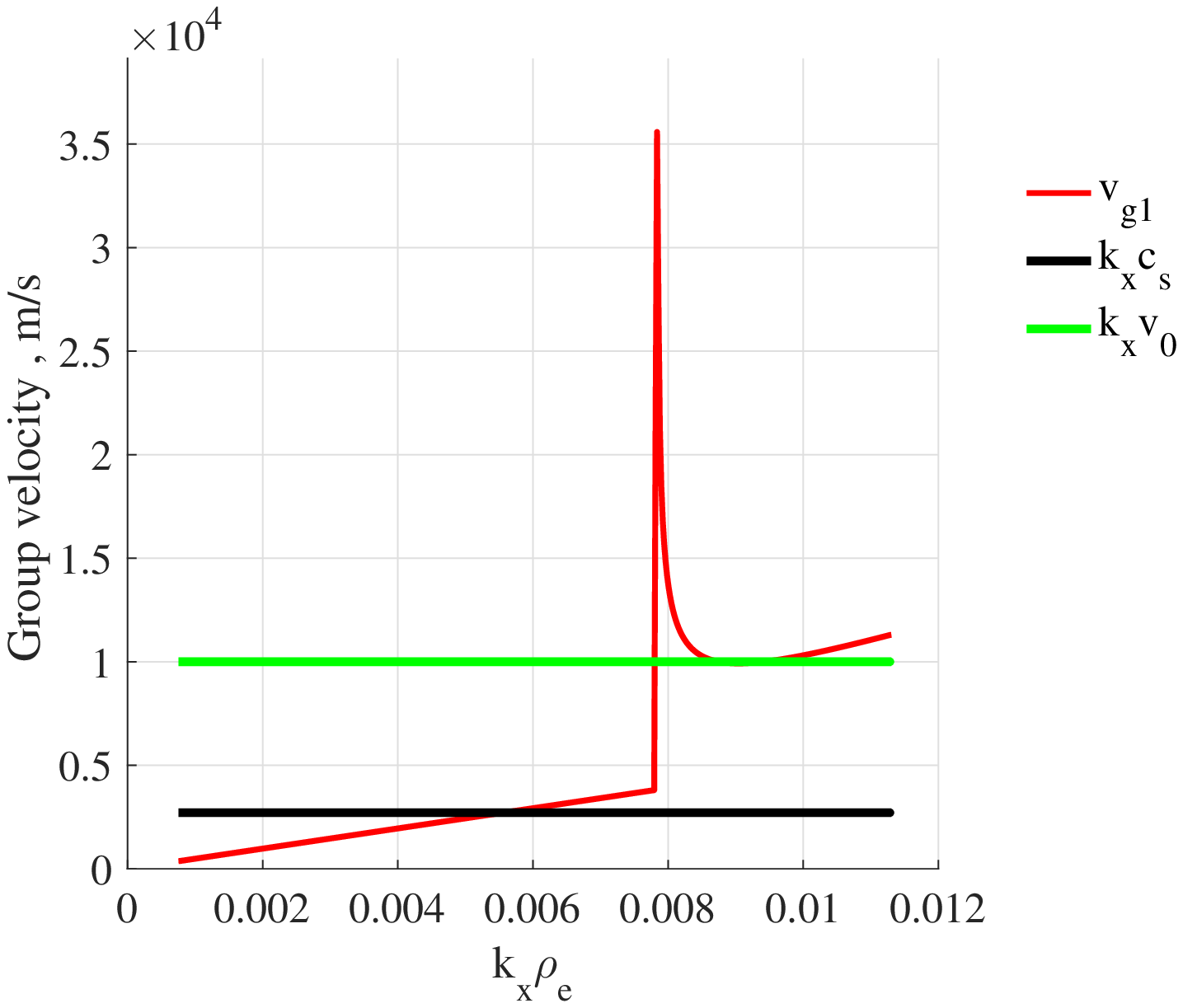}
	\caption{Group velocity in $y$ (left) and $x$ (right) directions.}
	\label{fig:inertiaGV}
\end{figure}

Group velocity was calculated as well, results are shown in Fig. \ref{fig:inertiaGV}. It is seen that in $y$ direction group velocity tend to be equal to $v_0$ velocity.

\begin{figure}[h!]
	\centering
	\includegraphics[width=8cm]{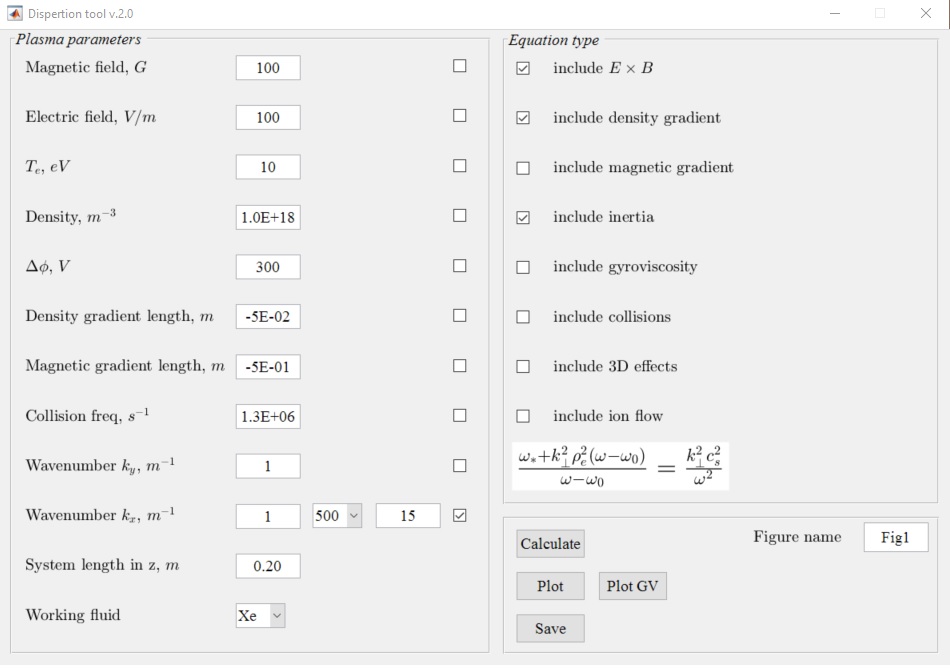}
	\caption{View of the solver interface with parameters, which were used to obtain Fig. \ref{fig:inertia}, \ref{fig:inertiaGV}.}
	\label{fig:interface_inertia}
\end{figure}

Solver interface with parameters, for Simon-Hoh instability with included inertia is shown in Fig. \ref{fig:interface_inertia}.

\subsection{Effects of electron inertia and gyroviscosity on Simon-Hoh modes}

In general, modes with high $k_{\bot }\rho _{e}\gg 1$ values should be considered with kinetic theory, due to importance of finite Larmor radius effect at such regimes. However, this effect can be incorporated into fluid consideration via the electron gyroviscosity \cite{SmolyakovPPCF2016}. Modified equation is shown below 
\begin{equation}
\dfrac{\omega _{\ast }+k_{\perp }^{2}\rho _{e}^{2}(\omega -\omega _{0})}{%
	\omega -\omega _{0}+k_{\perp }^{2}\rho _{e}^{2}(\omega -\omega _{0})}=\dfrac{%
	k_{\perp }^{2}c_{s}^{2}}{\omega ^{2}}.
\end{equation}

\begin{figure}[h!]
	\centering
	\includegraphics[width=6cm]{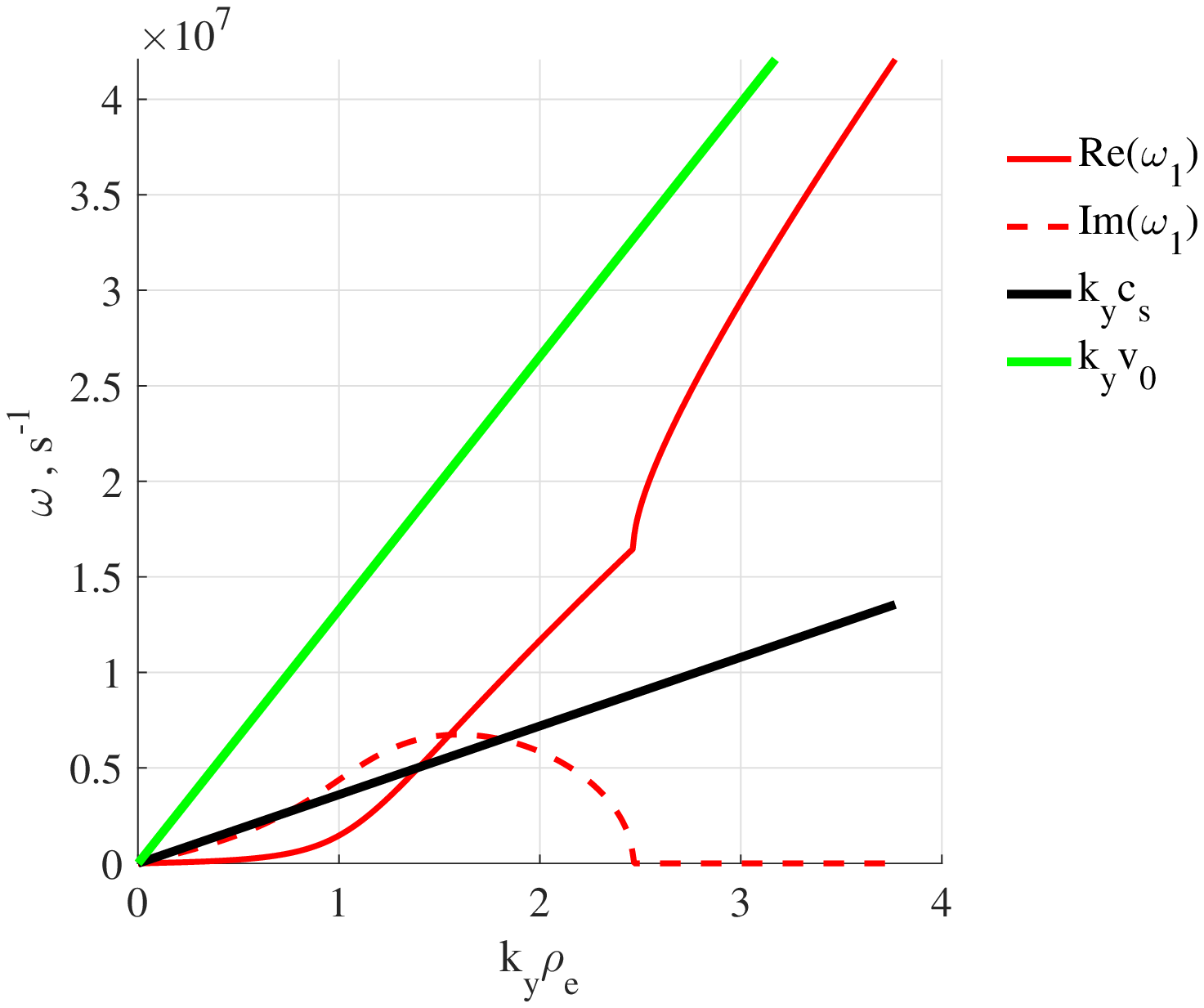}
	\includegraphics[width=6cm]{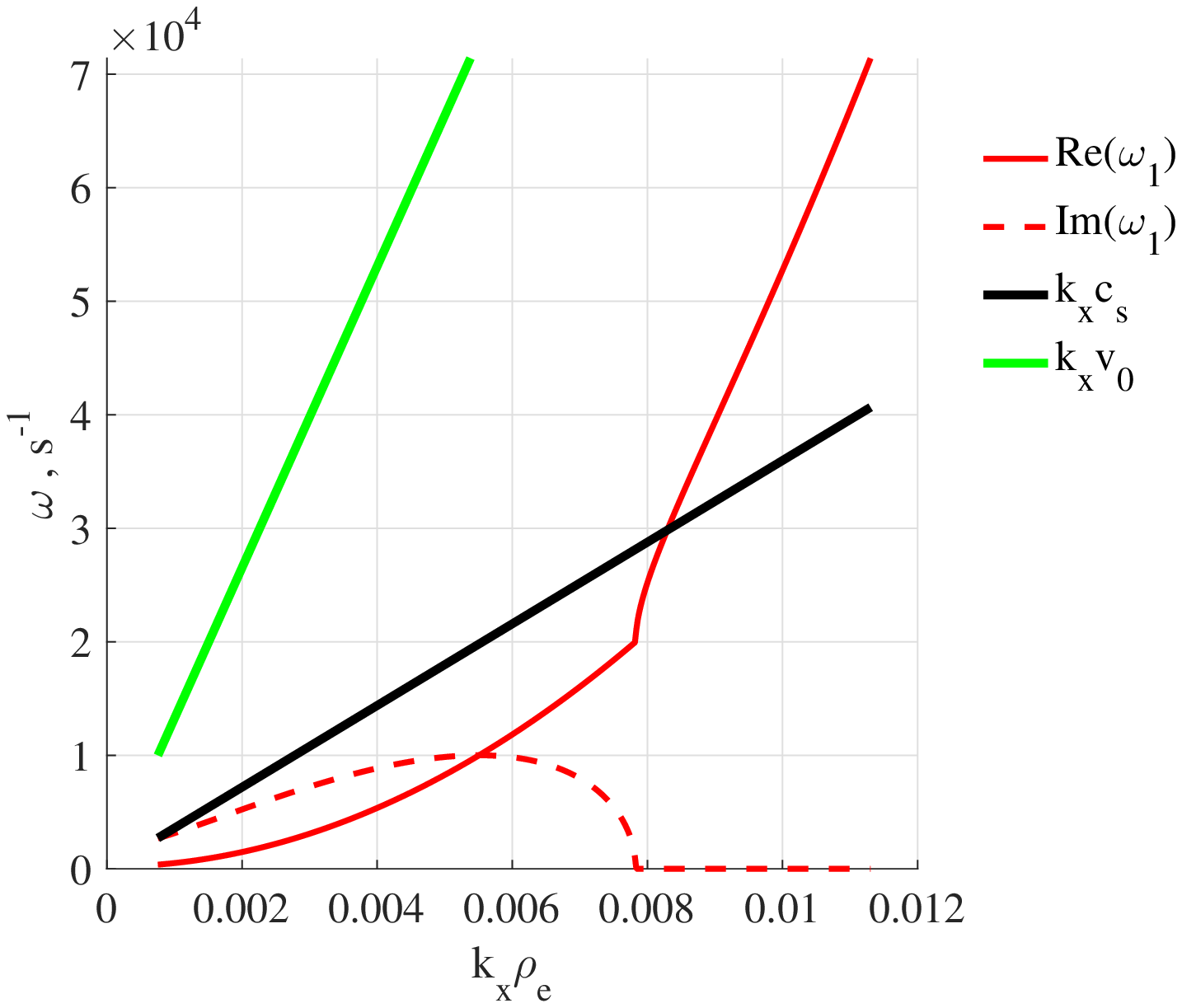}
		\caption{Frequency and growth rate of the Simon-Hoh instability with electron inertia an gyroviscosity effects as functions of $k_y\protect\rho_e$ at $k_x = 1$ (left) and $k_x\protect\rho_e$ at $k_y = 1$ (right).}
	\label{fig:inGV}
\end{figure}

Dependencies of unstable modes on $k_y\rho_e$ and $k_x\rho_e$ are shown in Fig. \ref{fig:inGV}. In general, gyroviscosity shifts unstable region towards higher values of $k_y$, and almost does not affect unstable region in $x$ direction.

\begin{figure}[h!]
	\centering
	\includegraphics[width=6cm]{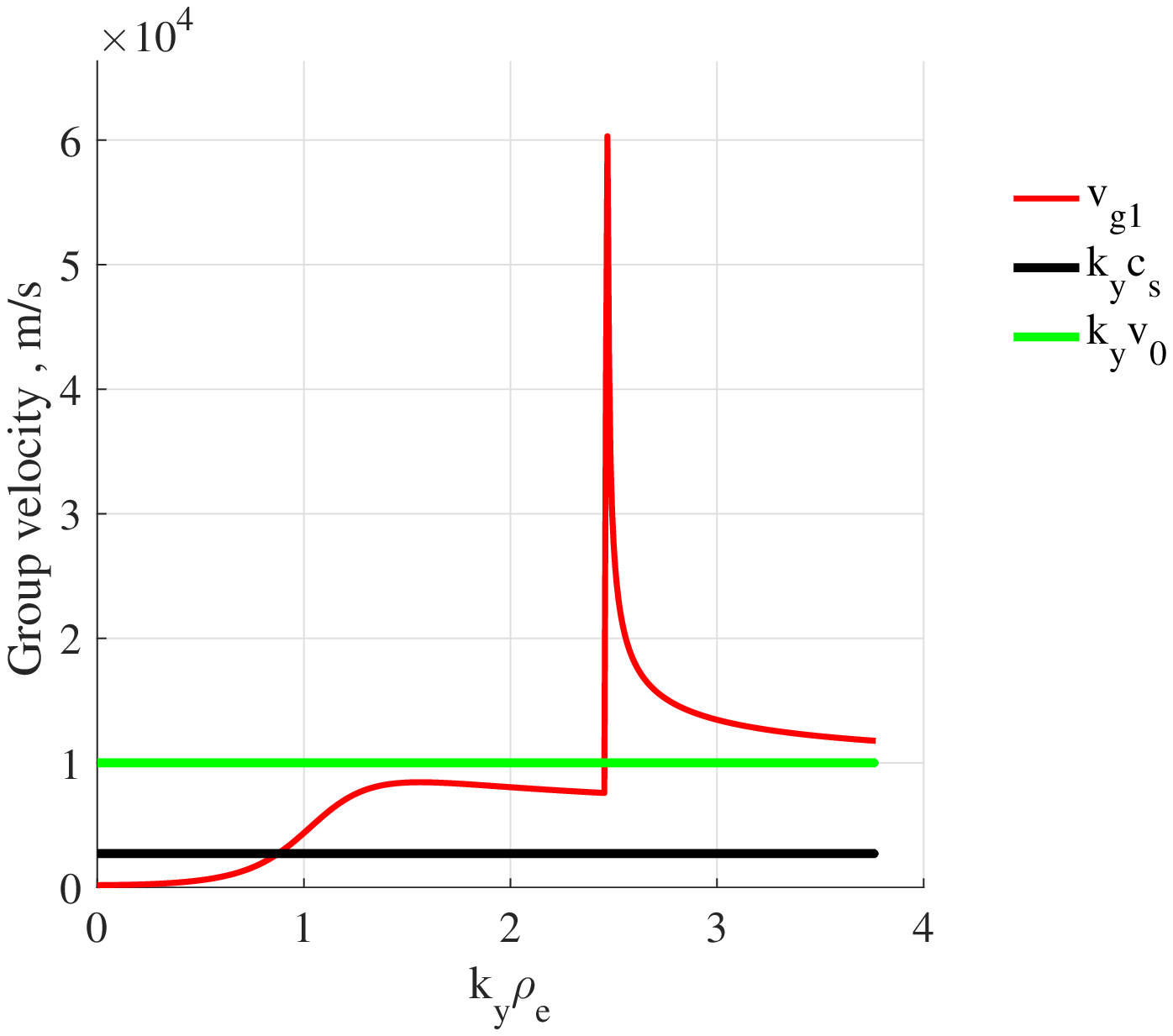}
	\includegraphics[width=6cm]{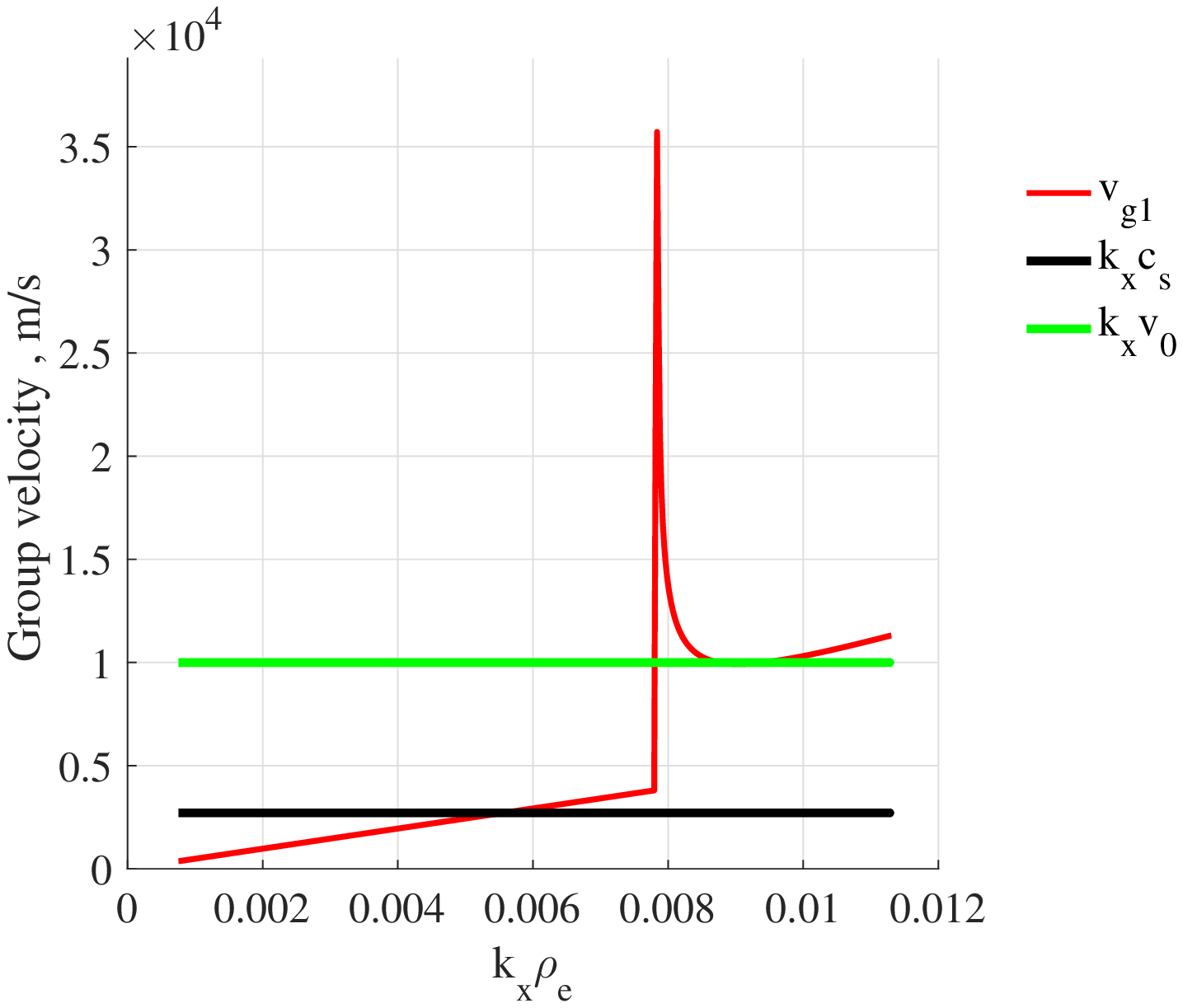}
	\caption{Group velocity in $y$ (left) and $x$ (right) directions.}
	\label{fig:inGV_GV}
\end{figure}

Group velocities in $x$ and $y$ directions are presented in Fig. \ref{fig:inGV_GV}. As before, maximum group velocity in $y$ direction is defined by $v_0$ velocity.

\begin{figure}[h!]
	\centering
	\includegraphics[width=8cm]{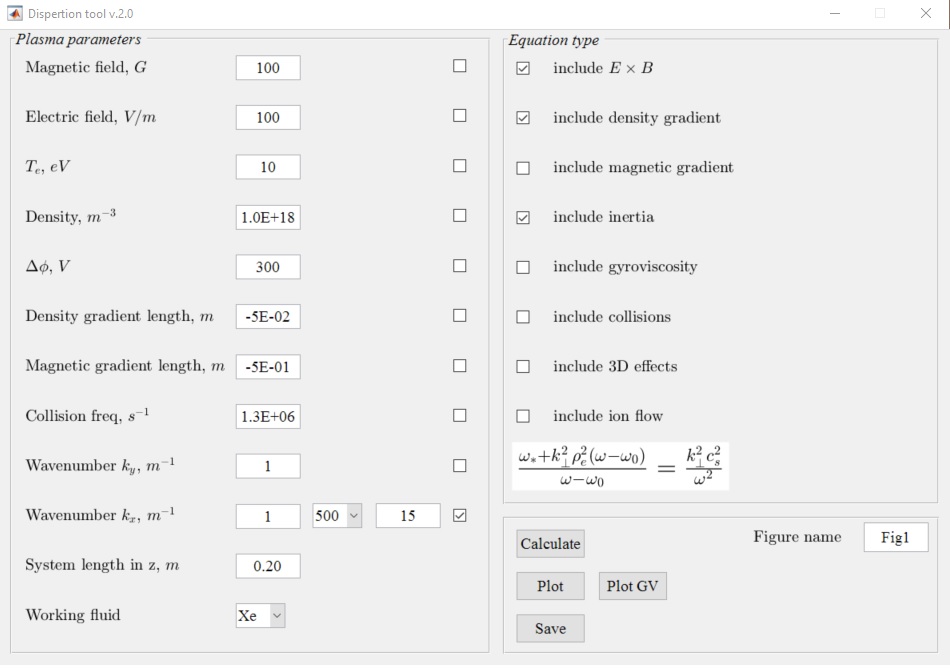}
	\caption{View of the solver interface with parameters, which were used to obtain Fig. \ref{fig:inGV}, \ref{fig:inGV_GV}.}
	\label{fig:interface_inGV}
\end{figure}

Solver interface is presented in Fig. \ref{fig:interface_inGV}.

\subsection{Full dispersion relation including effects of electron inertia, gyroviscosity, $\mathbf{E}\times \mathbf{B}$ flow, and collisions}

In systems with flows as typical for Hall discharge devices, there are modes with negative energy which become unstable due to the dissipation effects and collisions have been considered as an additional destabilization mechanism. The dispersion relation with effects of collisions has the form

\begin{equation}
\dfrac{\omega _{\ast }+k_{\perp }^{2}\rho _{e}^{2}(\omega -\omega _{0}+i\nu
	_{e})}{\omega -\omega _{0}+k_{\perp }^{2}\rho _{e}^{2}(\omega -\omega
	_{0}+i\nu _{e})}=\dfrac{k_{\perp }^{2}c_{s}^{2}}{\omega ^{2}}.
\end{equation}

\begin{figure}[h!]
	\centering
	\includegraphics[width=6cm]{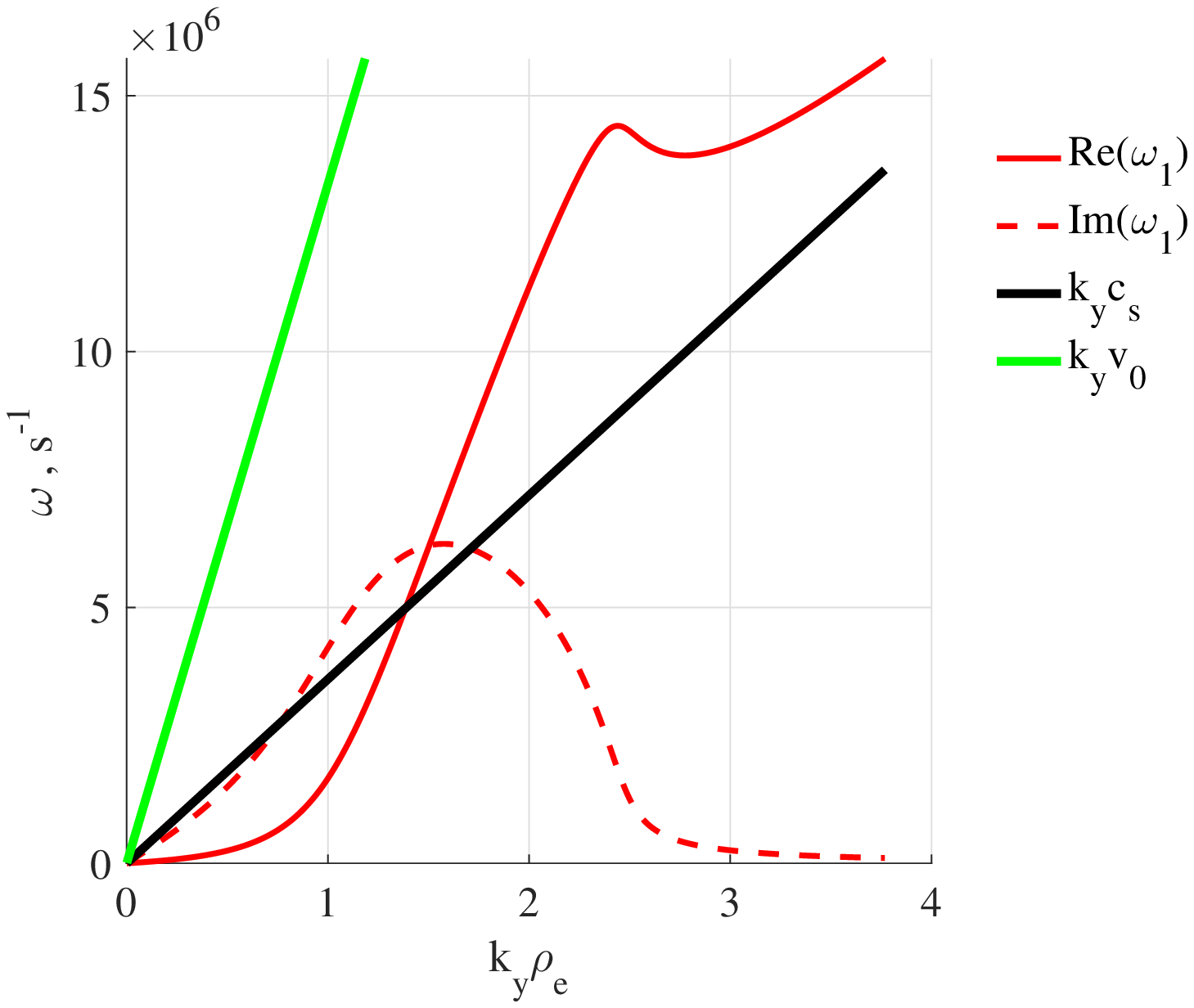}
	\includegraphics[width=6cm]{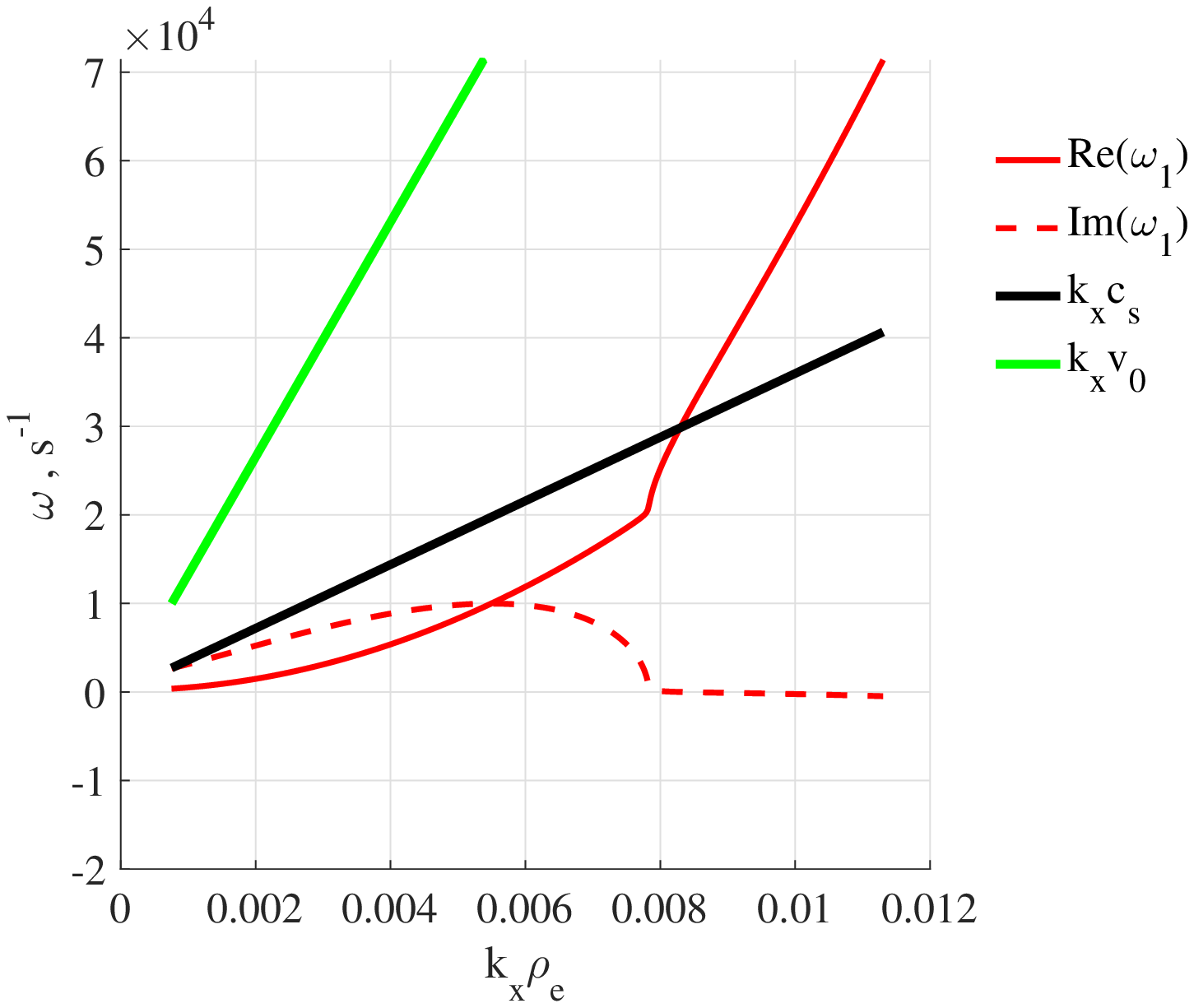}
	\caption{Frequency and growth rate of the Simon-Hoh instability with electron inertia an gyroviscosity effects as functions of $k_y\protect\rho_e$ at $k_x = 1$ (left) and $k_x\protect\rho_e$ at $k_y = 1$ (right).}
	\label{fig:coll}
\end{figure}

Frequencies and growth rates as functions of $k_y\rho_e$ and $k_x\rho_e$ are shown in Fig. \ref{fig:coll}. Collisions do not affect values of maximum growth rates and frequencies a lot; however, they extend unstable region to higher wavenumbers.

\begin{figure}[h!]
	\centering
	\includegraphics[width=6cm]{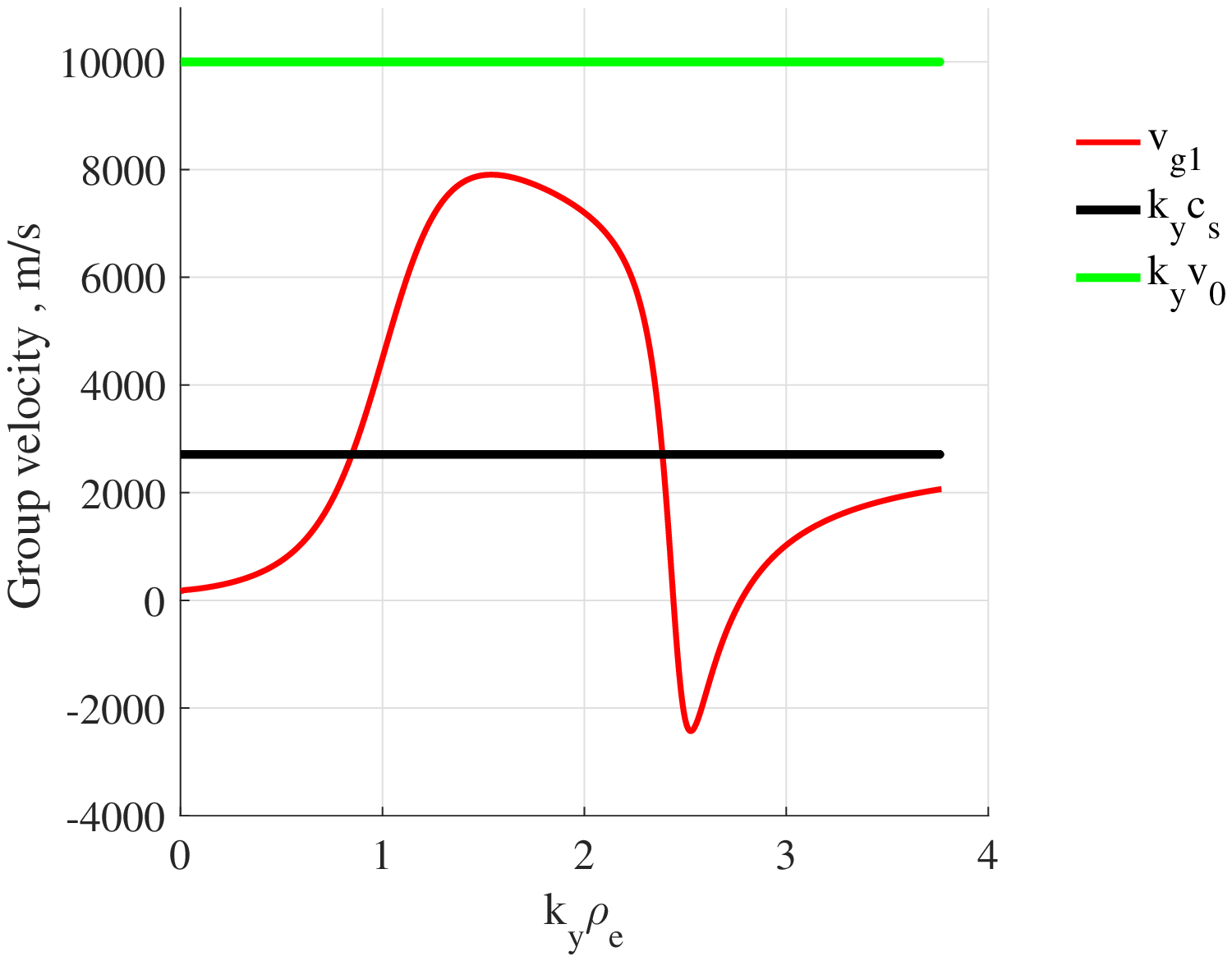}
	\includegraphics[width=6cm]{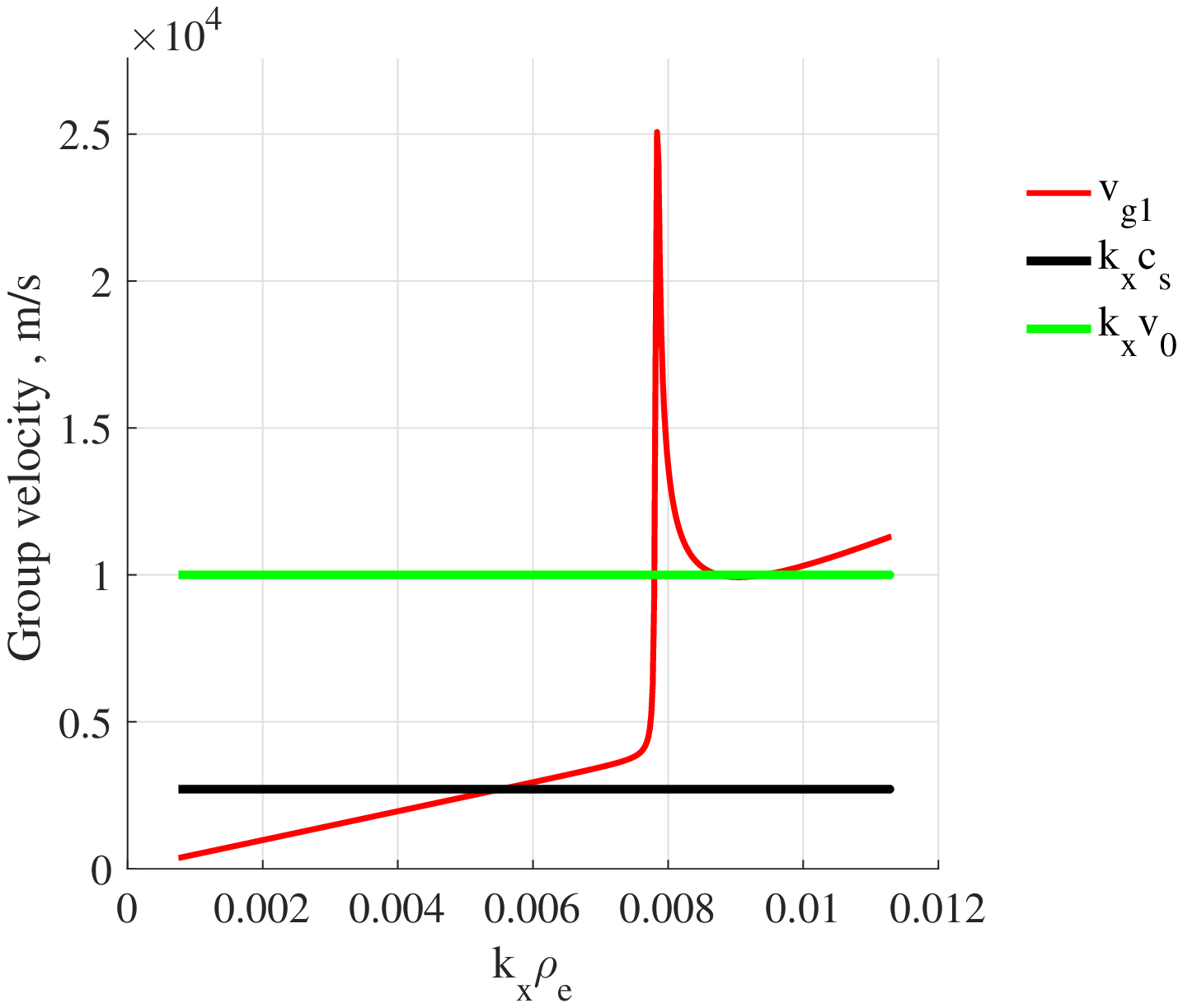}
	\caption{Group velocity in $y$ (left) and $x$ (right) directions.}
	\label{fig:coll_GV}
\end{figure}

Group velocities in $x$ and $y$ directions are presented in Fig. \ref{fig:coll_GV}. Due to more smooth function of $\omega(k_y)$, group velocity doe s not have an abrupt changes. The magnitude of group velocity is close to $v_0$.

\begin{figure}[h!]
	\centering
	\includegraphics[width=8cm]{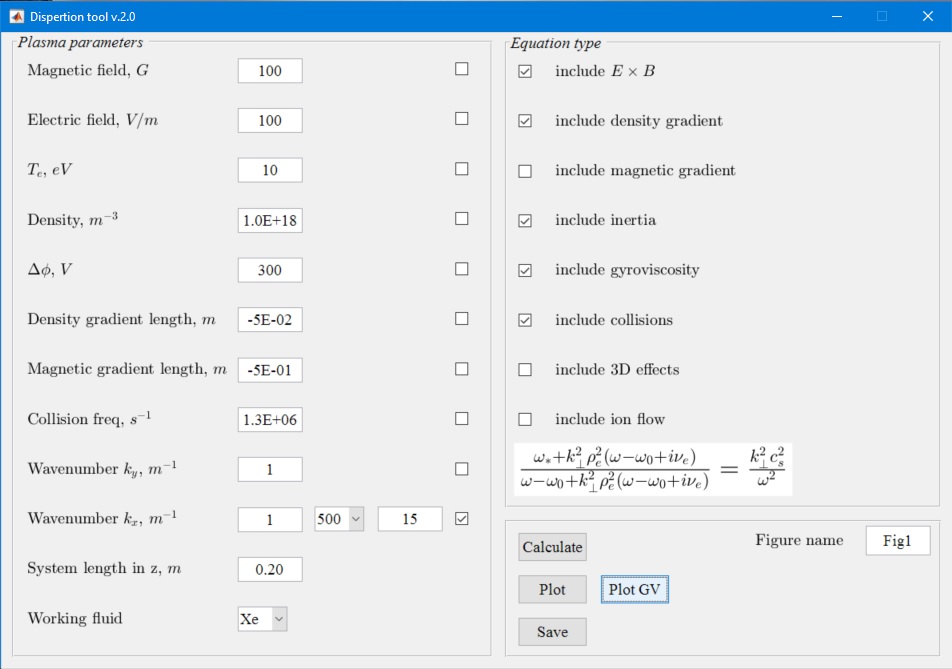}
	\caption{View of the solver interface with parameters, which were used to obtain Fig. \ref{fig:coll}, \ref{fig:coll_GV}.}
	\label{fig:interface_coll}
\end{figure}

Solver interface is presented in Fig. \ref{fig:interface_coll}.

\subsection{Effect of magnetic field gradient}

\begin{figure}[h!]
	\centering
	\includegraphics[width=6cm]{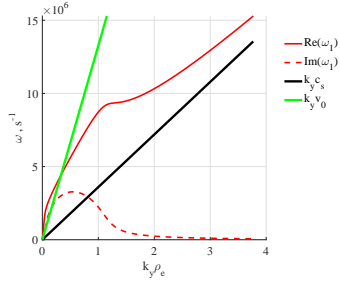}
	\caption{Frequency and growth rate from full dispersion relation with magnetic field gradient as functions of $k_y\protect\rho_e$ at $k_x = 1$.}
	\label{fig:Mag}
\end{figure}

There is no magnetic field gradients in Penning discharge, however this solver can be used with Hall thruster parameters, where magnetic field gradients play an important role. Thus, here we show the effect of the magnetic field gradient on Simon-Hoh modes and on full dispersion relation. Results are shown in Fig. \ref{fig:Mag}.

\newpage
Magnetic field gradient stabilizes the instability. For selected parameter, there is no instability for Simin-Hoh modes. Calculation for full dispersion relation shows that the magnitude of the maximum growth rate was significantly reduced. For the dependency on $k_x$ wavenumber it is too small and is not shown here.

\begin{figure}[h!]
	\centering
	\includegraphics[width=8cm]{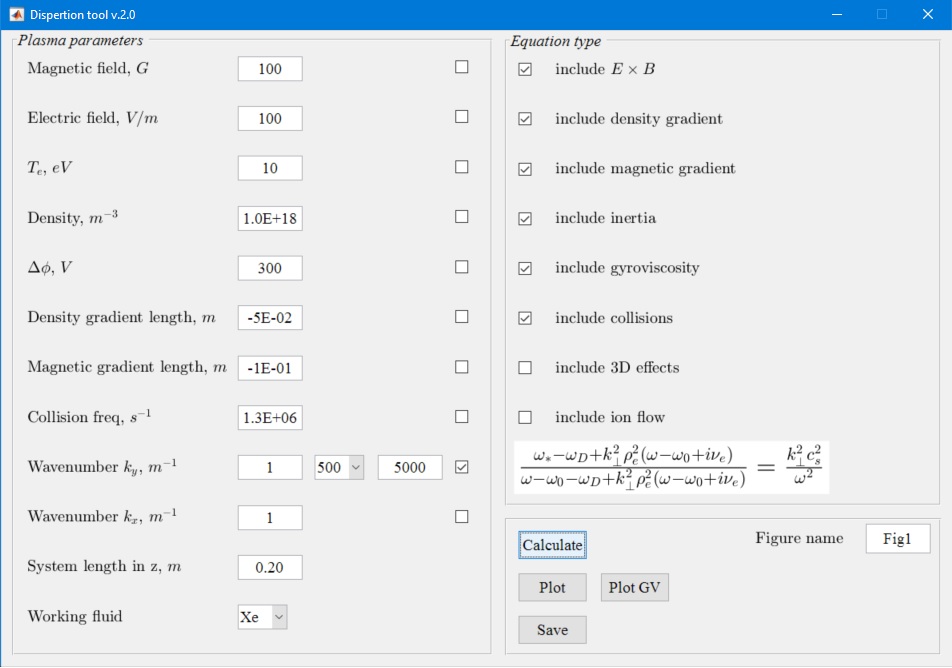}
	\caption{View of the solver interface with parameters, which were used to obtain Fig. \ref{fig:coll}.}
	\label{fig:interface_Mag}
\end{figure}

Solver interface is presented in Fig. \ref{fig:interface_Mag}.

\subsection{Search for growth rate and frequency as function of wavenumbers}

\begin{figure}[h!]
	\centering
	\includegraphics[width=6cm]{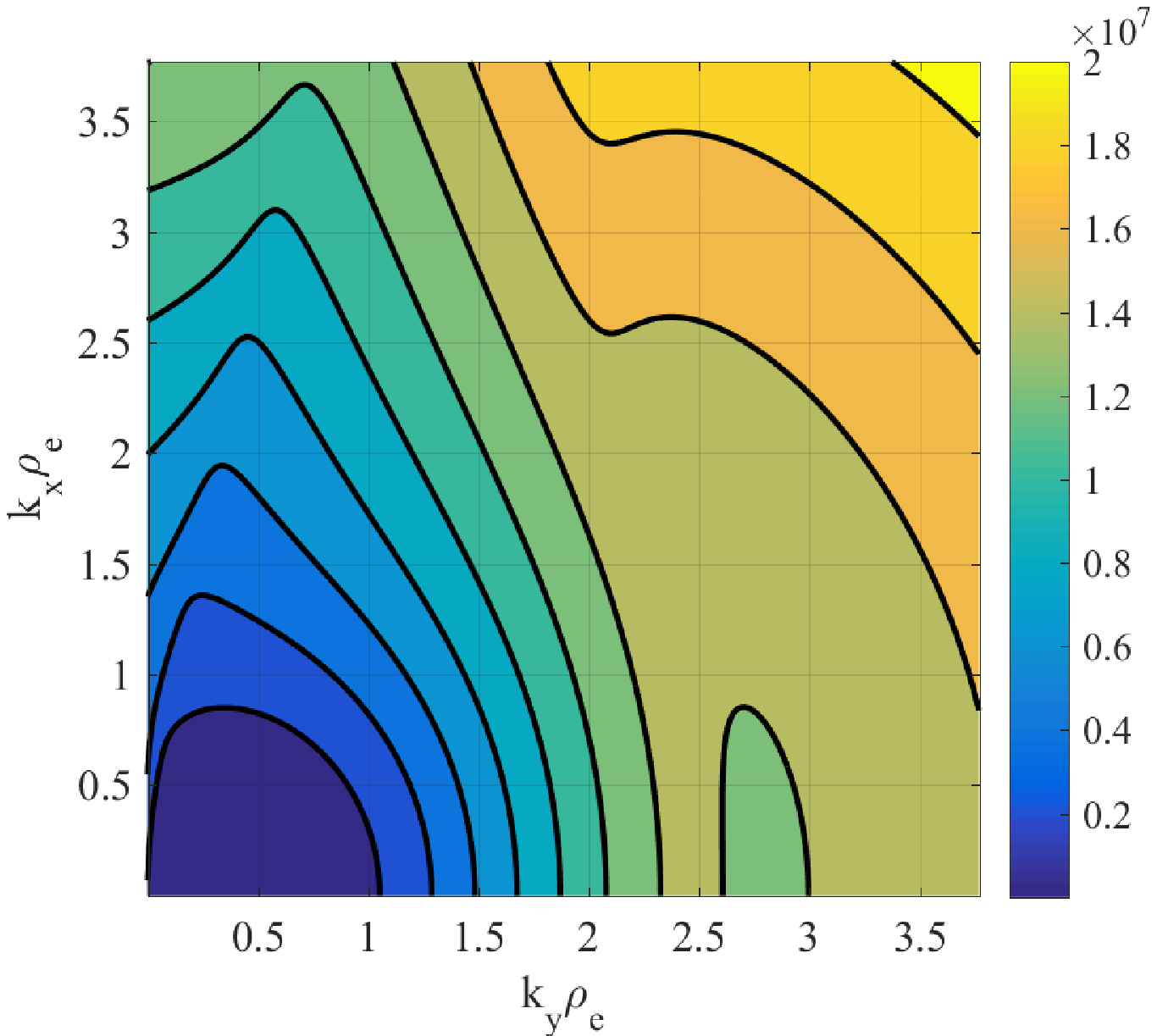}
	\includegraphics[width=6cm]{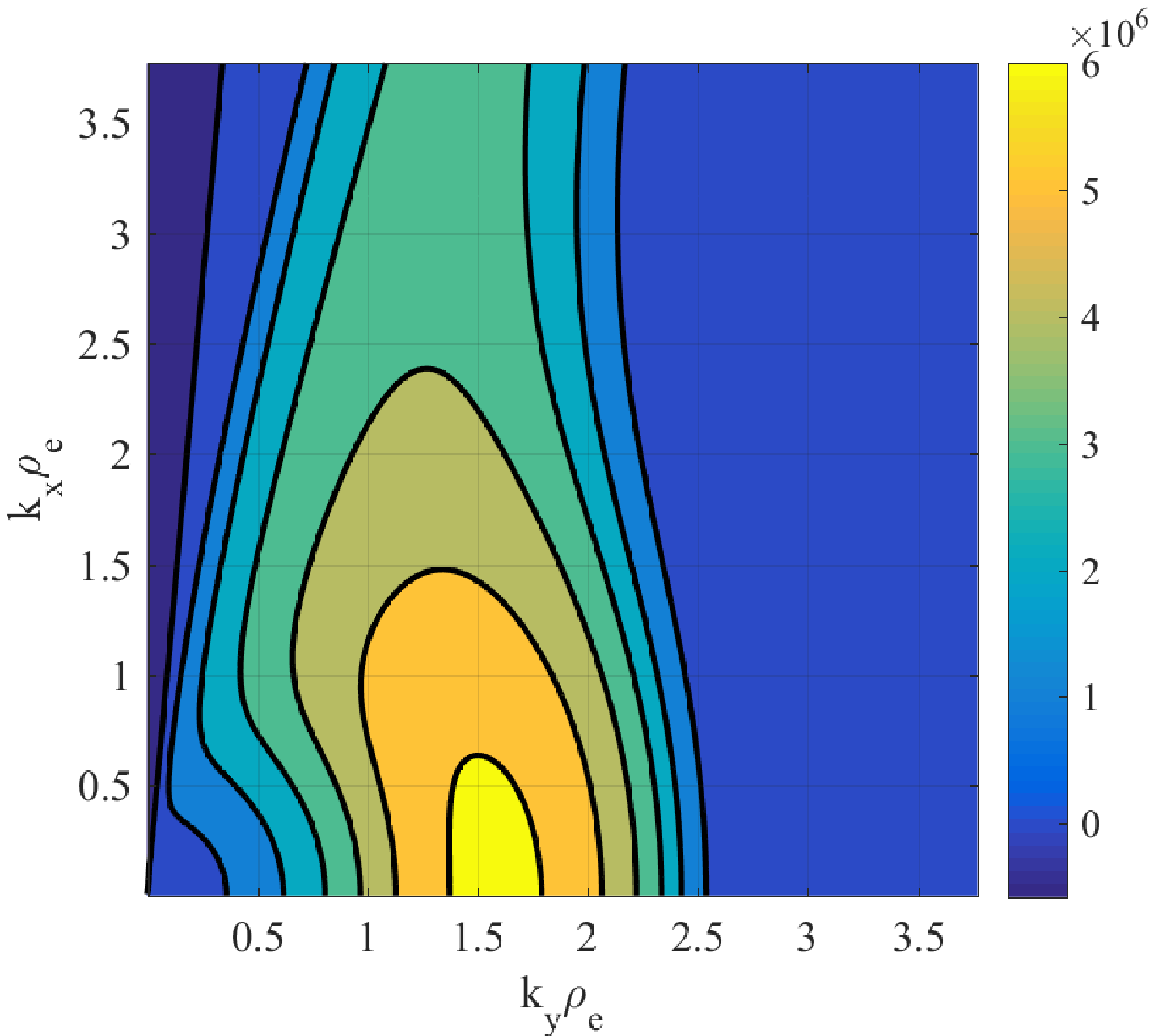}
	\caption{Frequency and growth rate.}
	\label{fig:3D}
\end{figure}

It is possible to plot a dependency of frequency and growth rate for unstable mode as a function of both $k_y\rho_e$ and $k_y\rho_e$. The results is two surfaces with represent frequency and growth rate. Example of such surfaces is given in Fig. \ref{fig:3D}.

\newpage
Group velocity can bu calculated as well. Result is shown in Fig. 

\begin{figure}[h!]
	\centering
	\includegraphics[width=6cm]{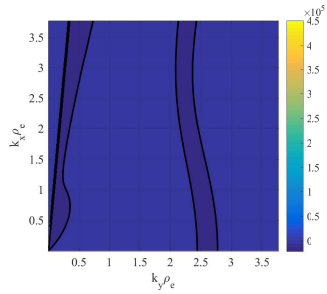}
	\caption{Group velocity.}
	\label{fig:3D_GV}
\end{figure}

\begin{figure}[h!]
	\centering
	\includegraphics[width=8cm]{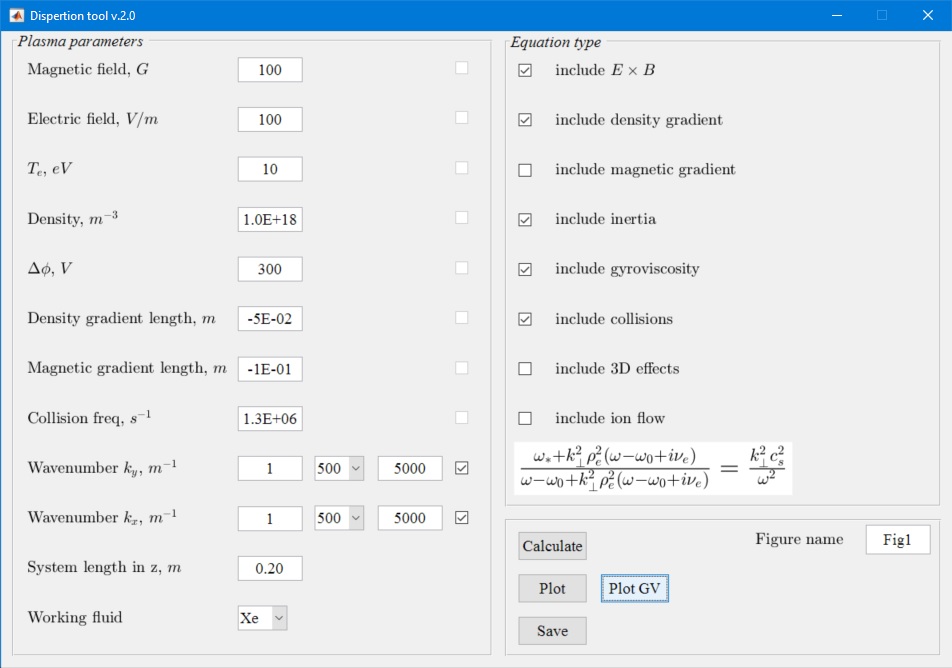}
	\caption{View of the solver interface with parameters, which were used to obtain Fig. \ref{fig:3D}.}
	\label{fig:interface_3D}
\end{figure}

Solver interface is presented in Fig. \ref{fig:interface_3D}.

\section{Conclusion}

Simple and convenient solver for general dispersion relation, which describes plasma in crossed $E$ and $B$ field with unmagnetized ions was developed. It allows to make estimates for frequencies and growth rates of the unstable modes under various plasma parameters. Such estimations can be used for comparison with fluid or kinetic simulations, or in real experiments.

Several cases of unstable modes with different physical mechanisms were considered. An interesting property of the local modes is that the real part of the frequency and growth rate of the most unstable mode are equal to the local $\mathbf{E\times B}$ frequency, $\omega =\omega _{0}+i\omega _{0}$ for arbitrary value of the $k_{y}$.


\begin{thebibliography}{99}
	\bibitem{SmolyakovPPCF2016} A.~Smolyakov, O.~Chapurin. W.~Frias,
	O.~Koshkarov, I.~Romadanov, T.~Tang, M.~Umansky, Y.~Raitses, I.~Kaganovich,
	V.~Lakhin, \newblock Plasma Physics and Controlled Fusion {\bf 59}, 014041
	(2017).
	\bibitem{Git} 
	\newblock {\url{https://bitbucket.org/ivr509/hall-plasmas-discharge-solver/downloads}}
	
	\bibitem{RaitsesIEPC2015} Y.~{Raitses}, I.~{Kaganovich}, and A.~{Smolyakov}, 
	\newblock {"Effects of the gas pressure on low frequency oscillations in
		{ExB} discharges}," \newblock in {\ Proceedings of Joint Conference of 30th 
		{\ ISTS, 34th IEPC and 6th NSAT}}, IEPC-2015-307.

	\bibitem{FriasPoP2013} W.~Frias, A.~Smolyakov, I.~D. Kaganovich, and
	Y.~Raitses, \newblock Physics of Plasmas {\bf 20}, 011527 (2013).
	
	\bibitem{FriasPoP2014} W.~Frias, A.~Smolyakov, I.~D. Kaganovich, and
	Y.~Raitses, \newblock Physics of Plasmas {\bf 21}, 042105 (2014).
	
	\bibitem{SimonPF1963}
	A.~Simon,
	\newblock Physics of Fluids {\bf 6}, 382 (1963).
	
	\bibitem{HohPF1963}
	F.~C. Hoh,
	\newblock Physics of Fluids {\bf 6}, 1184 (1963).
	
	\bibitem{TaoPF1993}
	Y.~Q. Tao, R.~W. Conn, L.~Schmitz, and G.~Tynan,
	\newblock Physics of Plasmas {\bf 1}, 3193 (1994).
	
	\bibitem{SakawaESHB}
	Y.~Sakawa, C.~Joshi, P.~K. Kaw, F.~F. Chen, and V.~K. Jain,
	\newblock Physics of Fluids B-Plasma Physics {\bf 5}, 1681 (1993).	
\end{thebibliography}
\end{document}